\documentclass[letterpaper]{JHEP}

\usepackage{epsfig}

\preprint{SUGP-99/12-1\\
 hep-th/9912184}

\title{Brane Transmutation in Supergravity}
\author{Andr\'es Gomberoff\\Physics Department, Syracuse University, Syracuse,
         New York 13244\\Centro de Estudios Cient\'{\i}ficos de Santiago, Casilla 
16433, Santiago 9, Chile}
\author{Donald Marolf\\Physics Department, Syracuse University, Syracuse,
         New York 13244}
\date{December, 1999}
\abstract{
We study a family of BPS solutions
of type IIA supergravity  that  can be interpreted as
describing the `transmutation' of a Neveu-Schwarz five-brane
into a D4-brane in the presence of a D6-brane.  The D4-brane,
which terminates on the D6-brane, can be equally well interpreted
as a `pure multipole' configuration of NS5-brane wrapped tightly
around the D6-brane.   Such a transmutation is a ``near-core''
version (i.e., near the D6-brane) of the brane-creation that can occur
when two branes pass through each other, as in the Hanany-Witten
construction.  The work below highlights certain 
charge non-conservation  features of type IIA supergravity.}
\begin{document}
\section{Introduction}

\label{intro}
It is by now a familiar story that string and M-theory contain various
types of branes (see \cite{Joe,KS,Youm,MAZ} for introductions and reviews)
which play a fundamental role in our understanding
of the theories.  In addition to elucidating the various dualities, 
they are central to the AdS/CFT correspondence
\cite{Juan,Magoo}
and its generalizations \cite{IMSY} and to matrix theory \cite{matrix}.
Typically, the various branes
are a net source of some gauge field, with a total charge that can
be measured by surrounding the brane with an appropriate 
Gauss' law surface and integrating either the field strength or its
dual. Measuring the various charges can tell us what sorts
of branes are present in a spacetime.

Now, in the associated supergravity theories, not all types of charge
are conserved in all cases.  As a result, in sufficiently
complicated situations, it is possible for one type of brane to
`transmute' into another.  
The family of type IIA supergravity solutions
investigated below turns out to just such a case, in
which the presence of a D6-brane catalyzes the transmutation of  
an NS5-brane into a D4-brane.  This is the supergravity
description near the D6-brane core 
of the sort of brane creation that arises in the Hanany-Witten construction
\cite{HW} and other examples \cite{Dan,Bachas,Bergman,Im,Cal} when two 
appropriate branes cross.  The family of exact supergravity solutions
presented below provides a supergravity moduli space in which
the dynamics of such processes might be further studied.

Interestingly, in our near-core family the NS5-brane seems to disappear
completely despite the fact that the associated current
{\it is} conserved in the setting considered below.
What happens is that the NS5-brane hides itself by folding into
a ``tightly wrapped'' configuration that produces no net monopole field.
The shape of the resulting 5-brane
can be thought of as the limit of a (2-dimensional)
paraboloid $P$ crossed with a flat 3-space ($P \times {\bf R}^3$) in
which the paraboloid degenerates to a half-line.
It turns out that the field produced by these NS5-branes
does not vanish.  However, as no Gaussian surface can
thread through the (zero-sized) opening of the paraboloid
to capture a net flux of field,
the net charge of such objects is zero for all practical purposes.
Thus, such brane configurations may be called `fundamental NS5
multipoles:'  the field they produce has no monopole part, but contains
only the higher
multipole moments\footnote{In curved spacetime, it
is not clear that there is a useful distinction between
dipole, quadrapole, etc. fields.  However, it is certainly
meaningful to refer to fields with zero monopole charge as `pure multipole
fields.'}.

Nonetheless, due to charge non-conservation effects in type IIA
supergravity, our  NS5-brane in fact carries
a nonzero net D4-brane charge.  In the limit in which the
paraboloid degenerates to a half-string,
the brane may equally well be interpreted as a D4-brane ending
on the D6-brane.  Now, the family of solutions considered below
forms a moduli space that
interpolates between a configuration consisting of a flat NS5-brane
widely separated from a D6-brane and the configuration with
a D4-brane ending on the D6-brane.  Thus, by the usual
adiabatic arguments, we may consider this family of solutions
to represent a {\it dynamical} process in which a slowly moving
NS5-brane approaches from infinity, wraps itself around a D6-brane, 
and transmutes into a D4-brane ending on the D6-brane.

The family of solutions to be studied below was in fact originally
constructed by Hashimoto in \cite{Aki}, using the method of
Itzhaki, Tseytlin, and Yankielowicz \cite{ITY}.
It is obtained by considering the `near-core'
solutions for
collections of M5-branes intersecting a Kaluza-Klein
monopole and performing a Kaluza-Klein reduction.  The result 
must be
a collection of D4- and NS5-branes intersecting a
D6-brane.  By taking into account certain
technical subtleties, we uncover the phenomena described above and
resolve the `puzzles' concerning such solutions that were
raised in \cite{Aki}.

Another version of the Hanny-Witten process was
previously studied \cite{Im,Cal}
using the Born-Infeld
worldvolume action to describe the creation of strings as a test D5-brane
wraps a D3-brane.
Although the back-reaction of the D5-brane on the metric is neglected
in such a treatment, the results for the shape of the D5-brane are
very similar to ours.  The original work \cite{Im} produced a description
near the D3-brane core and the authors of \cite{Cal} were able to
extend this to a BPS embedding of the D5-brane in the asymptotically
flat D3-brane background.  If our solutions could be extended to include
the asymptotically flat regions, it would be interesting to compare
the associated moduli spaces.

We begin the paper with a particularly transparent lower dimensional example 
in section \ref{ex}.  This example is closely related
to the one of central interest but several distracting features have
been removed.  The construction of the actual type IIA solutions
is reviewed in section \ref{review}, correcting few technical 
points that will be relevant to our discussion.  We then proceed in section
\ref{mapping} to map out the relevant branes, verifying that all
proceeds in parallel with the lower dimensional case.
We conclude with a short
discussion in section \ref{disc}.

\section{A transparent example}
\label{ex}

The set of solutions on which we will focus in sections \ref{review} and 
\ref{mapping} will be constructed following \cite{Aki}, and
using the method of \cite{ITY}, by first considering 
an M5-brane in flat spacetime.  Now, since the unit-charged Kaluza-Klein
monopole is in fact smooth at the center and since one expects that there
is in fact a solution\footnote{The intersection manifold is of sufficiently high dimension, see
\cite{amanda,dj}.} representing an M5-brane intersecting an asymptotically
flat charge $N$ monopole, the M5-brane in flat space should give the
``near-core'' version of a solution
representing an M5-brane intersecting a unit charged Kaluza-Klein
monopole.    By the appropriate $\bf Z_N$ quotient of this solution, 
one arrives at an M5-brane intersecting an ALE space which represents the
near-core solution for a  charge $N$ monopole.  
Kaluza-Klein reduction will then yield the near-core solution for a
collection of D4- and
NS5-branes intersecting a D6-brane.  

In the current section, we consider the simpler case of a membrane
in flat 4+1 Minkowski space, and then reduce it in much the same
manner as for the M5-brane mentioned above.  The result is a static
3+1 spacetime, for which the three spatial dimensions are easily visualized.
The D6-brane becomes simply a point 0-brane at the origin of the three
space.  The Kaluza-Klein reduction of the membrane yields a
collection of string and membrane charges in the 3+1 spacetime.
Such a lower dimensional example
sets the stage well for our discussion in sections \ref{review}
and \ref{mapping} and provides an opportunity to introduce
some notation. 

The example of the present section
is not intended to relate to any particular version of
4+1 supergravity, though it could certainly be made to do so by considering
4+1 Minkowski space as some compactification of 11-dimensional supergravity.
We simply take the membrane
to be a magnetic source (in analogy with the M5-brane to be considered below)
of some linear U(1) 1-form field strength $F^{4+1}_{[1]}$. 
The superscript on the field
strength refers to the fact that it represents a 4+1 field, as opposed
to the 3+1 fields that will arise in the reduction below.

For maximal compatibility with our higher dimensional discussion to come and
to fit with the notation of \cite{Aki}, let us refer to the four
Cartesian spatial coordinates as $x_7, x_8, x_9,$ and $x_{10}$.
These will be grouped into complex combinations:
\begin{equation}
W =  x_7 + i x_{10},  \ \ \ \ \ \ \ V =  x_8 + i x_9.
\end{equation}
We will use $w$ to denote the magnitude $|W|$ of $W$, and similarly
set $v=|V|$.  

We begin with a single two-brane at $V=0$.  Since this is a magnetic source 
in 4+1 dimensions, 
its total charge is measured by any 1-sphere which encloses the brane and
must be independent of the choice of this 1-sphere.  As a result, the
symmetries and the Bianchi identity $dF^{4+1}=0$ for $V \neq 0$ determine
the field strength to be of the form
\begin{equation}
F^{4+1}_{[1]} = \frac{Q}{2 \pi \left( 
x_8^2 + x_9^2\right)} \left(x_8 dx_{9} - x_{9} dx_8 \right),  
\end{equation}
where $Q$ denotes the total charge of the brane.
  
Let us now introduce three new coordinates ($\rho$, $\theta$, and
$\psi$) which will describe the reduced spacetime, and a coordinate
$\phi$ which will label points along the Killing orbits.  That is, in
the coordinates below, we will reduce along the Killing field
$k = \partial_\phi$.  The particular form of the coordinates is given by
\begin{eqnarray}
V &=& \rho \sin \frac{\theta}{2} \ e^{i(\psi + \phi)} \cr
W &=& \rho \cos \frac{\theta}{2} \ e^{i\phi}. \label{VWtransf}
\end{eqnarray}

Note that the Killing field $k$ acts as a simultaneous rotation of both the
$V$ and $W$ planes.   As a result, the only fixed point is at the
origin $V=W=0.$  It may be checked that this is the only singularity in the
reduction process.  Away from $V=W=0$, the orbits of the Killing field are
circles and smoothly foliate the 4+1 spacetime.  
In the type IIA case of sections \ref{review} and \ref{mapping},
when the gravitational
field of the M5-brane is ignored, $\theta$ and $\phi$ become the usual
angular coordinates of the D6-brane metric, and the metric reduces to the
that of the near-core D6-brane.  The D6-brane metric is of course smooth
outside the origin.  See \cite{IMSY,Aki,ITY} for details.  
Thus, we may think of $\rho, \theta, \phi$ as 
forming a standard set of spherical coordinates on ${\bf R}^3$, with
some singularity in the origin.  In particular, $\theta$ ranges over
$[0,\pi]$, with $\theta=0$ and $\theta=\pi$ representing the poles
of the 2-sphere.  In the current simplified example, this
singularity is point--like (a 0-brane).   This can also be seen from the
4--dimensional spatial metric
\begin{equation}
|dV|^2+|dW|^2 = d\rho^2 + \frac{\rho^2}{4}(d\theta^2+\sin^2\theta d\psi^2)
+ \rho^2
\left(d\phi +\sin^{2}\frac{\theta}{2} d\psi\right)^2 \ ,
\label{reducedALE}
\end{equation}
where  the first three terms define the ${\bf R}^3$ described above, and the
anomalous $1/4$
factor in front of the  2--sphere metric  gives rise to a singularity at
$\rho=0$.

Let us now study the orbits of $\partial_\phi$ that lie within the membrane.
We see that these are at $\theta =0$ and that there is one
orbit for each $\rho > 0$.
Note that each orbit lies at a coordinate singularity of $\psi$, and that
this collection of orbits projects
under the reduction to the $\theta =0$ axis.  As a result, the
projection of the membrane to the reduced spacetime
yields a string-shaped charge which terminates at the 
0-brane at the origin.

Here we reach a `puzzle,' related to the ones mentioned in
\cite{Aki}.  One is tempted to interpret the source along the $\theta = \pi$
axis as the usual charged string of
the 3+1 theory.  However, this leads to an immediate question about charge
conservation. In 3+1 dimensions, one measures the charge of a (magnetic)
string
by integrating a one-form field strength $F_{[1]}$ around a circle
enclosing the string.  If we consider such a circle that encloses
the string, we see that we can easily deform this circle to a point
by pulling it down past the origin and shrinking the circle to a point
on the negative
axis (see Fig. 1).

\EPSFIGURE{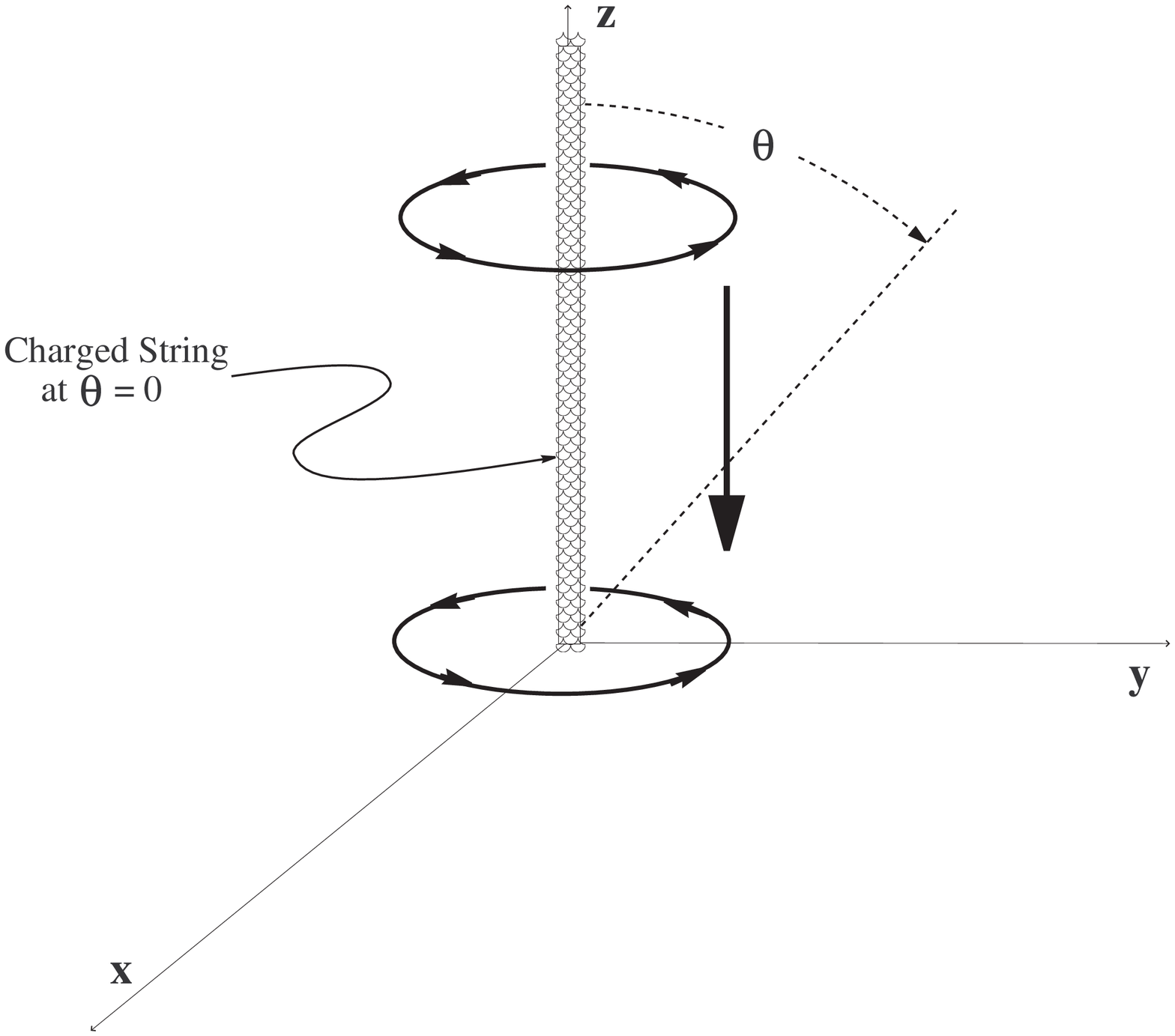,width=10cm}{Measuring the charge of the string.}

A second such puzzle is that Kaluza-Klein reduction of the 4+1 field strength
$F^{4+1}_{[1]}$ yields
both a 3+1 one-form field strength $F_{[1]}$ and a 3+1 
zero-form field strength $F_{[0]}$.  The reduction proceeds as
\begin{equation}
\label{Fdecomp}
F_{[1]}^{4+1} = F_{[1]} + F_{[0]} \wedge d\phi.
\end{equation}
For simplicity of notation, we use symbols without superscripts
to refer to the 3+1 dimensional fields.
In the present case this yields $F_{[1]} = \frac{Q}{2\pi} d \psi$, $F_{[0]} =
\frac{Q}{2\pi}$.  The puzzle is that the
zero-form field strength $F_{[0]}$ is associated with magnetic {\it membranes}
in 3+1 dimensions, although none seem to be present here.

For completeness, we mention that there also
 arises a 2-form field strength $F_{[2]}$ and an associated
vector potential $A_{[1]}$ from the Kaluza-Klein reduction of the
4+1 metric.  It turns out that $A_{[1]}$ describes
a magnetic monopole:
\begin{equation}
A_{[1]} = \frac{1}{2} (1 - \cos \theta) d \psi.
\end{equation}

Let us now return to the issue of charge conservation.
Since $F_{[1]}^{4+1}$ is a closed form, it follows from 
(\ref{Fdecomp}) that $F_{[1]}$ and
$F_{[0]}$ must be closed as well.  As a result, string charge 
defined by integrating $F_{[1]}$ around closed curves
should be conserved.  That this is in fact the case
is clear from the result $F_{[1]}= \frac{Q}{2\pi}  d \psi$.  Such a field
appears to refer to a string that enters from infinity along
the $\theta=0$ axis, passes through the origin, and then
exits to infinity along the $\theta = \pi$ axis.

However, 
a relevant issue is that the field strength $F_{[1]}$ is not
gauge invariant.  As we can see from (\ref{Fdecomp}), 
under a change of coordinates $\phi \rightarrow \phi + \gamma(\rho, 
\theta, \psi)$  (which is just a U(1) gauge transformation of the
vector potential $A_{1}$ that arises from the Kaluza-Klein reduction), 
we have $F_{[1]} \rightarrow F_{[1]} + F_{[0]} \wedge d \gamma$.
As a result, $F_{[1]}$ is a very subtle object in the presence of a
non-trivial $A_{[1]}$ bundle, such as that which describes our magnetic
monopole.  For this reason, one typically adds a term to $F_{[1]}$ to
make a gauge invariant field strength, defining
\begin{equation}
\tilde F_{[1]} = F_{[1]} - F_{[0]} \wedge A_{[1]}.
\end{equation}
Note that this field satisfies
\begin{equation}
d \tilde F_{[1]} + F_{[0]} \wedge F_{[2]} =0,
\end{equation}
in vacuum
since $F_{[0]}$ and $F_{[1]}$ are closed.

The importance of all this for our discussion becomes
clear when we couple a magnetic 
current $j_1$ (representing a string source).
The above equation of motion becomes:
\begin{equation}
\label{charge}
d \tilde F_{[1]} + F_{[0]} \wedge F_{[2]} ={}^*\! j_1,
\end{equation}
where $*$ represents the Hodge dual.
By taking the exterior derivative of this equation, we see
that the string current is in fact not conserved\footnote{Alternatively, 
one could define $*j_1 = d \tilde F$.  Such a current
is conserved, but is non-zero (and equal to $-F_{[0]} \wedge F_{[2]}$) 
even when no branes are present.  We prefer the definition 
(\ref{charge}) as it leads to the usual definition of brane 
charge \cite{Page,KS,GMT}.}.
We have
\begin{equation}
\label{cnc}
d{}^*\! j_1 = {}^*\! j_2 \wedge F_{[2]} + F_{[0]} \wedge {}^* \! j_0,
\end{equation}
where $j_2 = {}^*\!  dF_{[0]}$ is the current associated with membranes and 
$j_0 = {}^*\! dF_{[2]}$
is the current associated with the monopole at the origin.

Note that conservation cannot be restored by recognizing that, in
analogy with the case for M-branes, D-branes, and fundamental
strings \cite{CHS,T1},
certain fields may live on the the monopole and carry the charge
of the string.  As in \cite{witten}, one finds that the pullback
of certain bulk fields act as sources and sinks of such brane fields, so
that charge is still not conserved.

As a result, we see that the string current need not be
conserved when the other currents are nonzero.  In particular, it
is the presence of the monopole ($j_0$) together with the nonzero
$F_{[0]}$ field that allows the string to end at the origin.
In contrast, the membrane current $j_2$ is in fact
conserved in this model.

There remains, however, the issue of the nonzero $F_{[0]}$
field.  From whence does it arise?  We will see that it
arises because the string charge along the $\theta =0$
axis may equally well be interpreted as a tightly rolled
paraboloid of membrane.  It turns out that such a tightly
rolled membrane is in fact physically equivalent to a half-string.

To see that this is the case, we again follow the lead of 
\cite{Aki} and consider the more general family of solutions
obtained by placing not a single membrane at $V=0$ in the 4+1 spacetime,
but in fact
a uniform density of branes around the circle $|V| = b$.
The branes are still oriented along the $W$ plane, so that
we recover the configuration above in the limit $b \rightarrow 0$.
In this case we have
\begin{eqnarray}
F^{4+1}_{[1]} &=& \frac{Q}{2\pi} \left(\frac{x_8 dx_9 - x_9 dx_8}{x_8^2 + x_9^2}\right)
\ \ \ {\rm for} \ |V| > b, \\
F^{4+1}_{[1]} &=& 0
\hspace{4.1cm} {\rm for} \ |V| < b.
\end{eqnarray}
Thus, for $v = \rho \sin \frac{\theta}{2} > b$, the gauge fields
are independent of $b$, while they vanish\footnote{Except of course
$F_{[2]}$, which arises purely from the Kaluza-Klein reduction of
the 5-dimensional metric.} for $v < b$.  For the reader's convenience, 
a diagram showing the constant $v$ surfaces in the reduced
3-space is shown below in Fig.2.


\begin{center}
 \epsfxsize=10cm
 \leavevmode
 \epsfbox{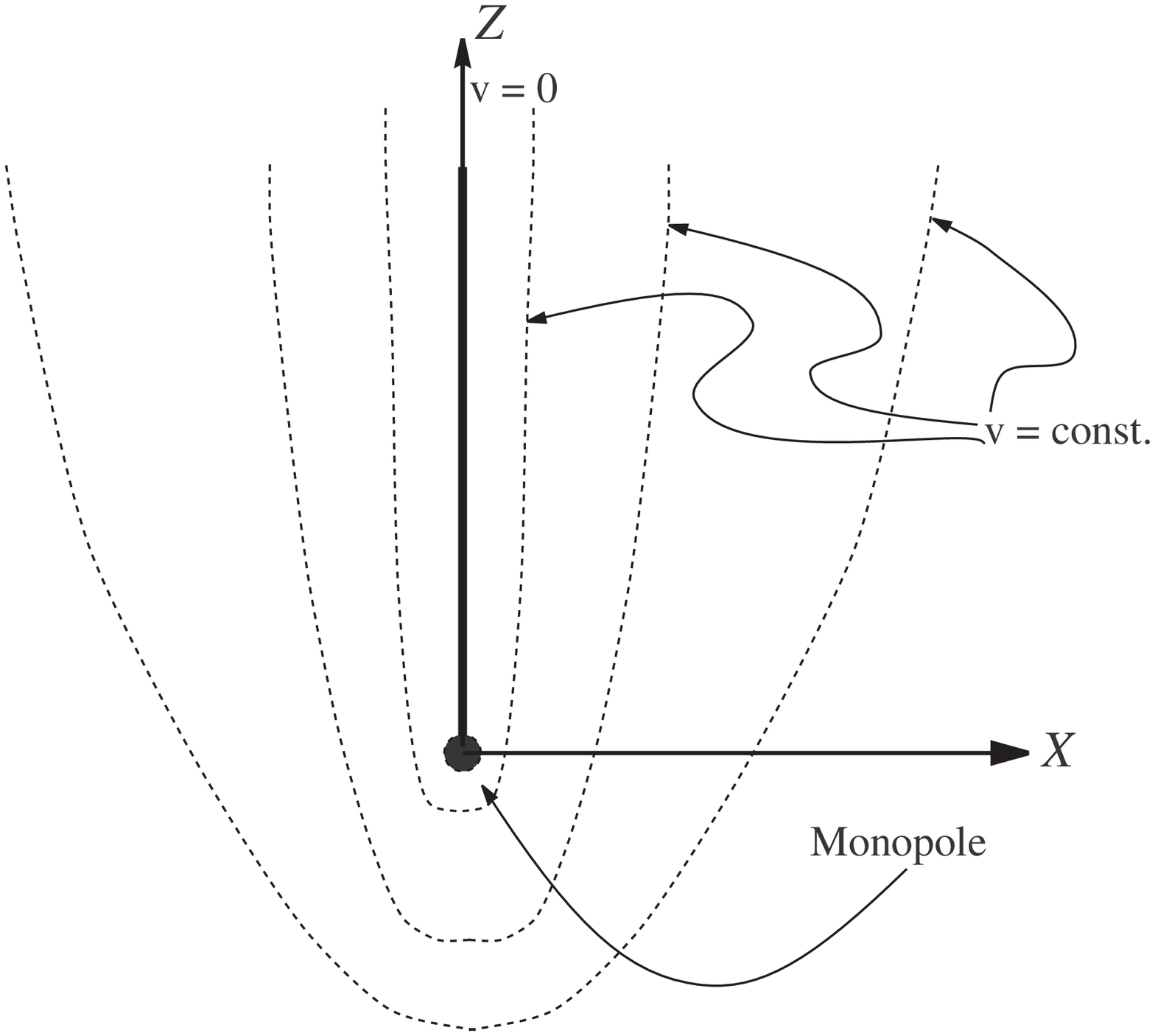}
\end{center}
\nopagebreak
{\small {\bf Figure 2}: The constant $v$ surfaces. The complete $3$--dimensional
reduced ALE space is obtained by rotating this planar diagram about the $Z$--axis.}

\noindent
For later reference we also  give the  explicit  projection from the $4$
dimensional space to the $3$--dimensional space depicted above:
\begin{eqnarray}
X &=& \bar{W}V + W\bar{V} =  R\sin\theta \cos\psi  \\
Y&=&  i(\bar{W}V - W\bar{V}) =  R\sin\theta \sin\psi  \\ 
Z &=&|W|^2-|V|^2=  R \cos\theta \ \ ,
\end{eqnarray}
where $R=\rho^2/2$.

We see that there is a discontinuity in $F_{[0]}$ across
the surface $v=b$.  Such a domain wall is associated with a membrane
charge sitting on the paraboloid $v=b$.  We may say that the membrane
generates this flux of $F_{[0]}$.  What is unusual about this
case is that the field strength $F_{[0]}$ does not
vanish in the $b \rightarrow 0$ limit in which the paraboloid
degenerates.   Instead, the presence of the monopole at the 
focus of the paraboloid stabilizes the $F_{[0]}$ field.
However, in the $b=0$ limit, one cannot see inside the degenerate
paraboloid and there are no longer any discontinuities
in the field strength.  Thus, the limiting distribution of branes
cannot be said to contain a net membrane charge.  Instead, we
refer to it as a `pure multipole' configuration of branes.
We will see that much the same thing occurs in our higher dimensional
case below, where the resulting field strength is more interesting.

So, for $b \neq 0$ it is clear that we have a membrane
present on the paraboloid.  What about the strings in this case?
Note that, since the membrane curves around the monopole, 
the term ${}^*\!  j_2 \wedge F_{[2]}$ is nonzero.  Thus, as noted
above, strings will not be conserved in this spacetime.  Instead, 
the membrane will act as a source of strings\footnote{On the other
hand, since the monopole is inside the paraboloid, we
have ${}^*\!  j_0 \wedge F_{[0]}=0$ and the 0-brane no longer
creates or destroys strings.}.   

Whether one wishes to
consider such strings as `separate' from the membrane or as
a part of the same object is simply a matter of record keeping.
The interesting question is whether we may consider {\it all}
strings in the spacetime to be generated by the membrane.
The answer to this question is `yes.'  This fact follows
from charge (non-)conservation.  We simply note that any strings
entering the spacetime must do so along the paraboloid in a
rotationally symmetric way, converging toward the vertex
of the paraboloid.  Thus, if charge were conserved, there
would need to be some net flux of strings exiting the vertex
of the paraboloid.  However, this is not the case.  All brane
sources in the spacetime are confined to the paraboloid 
itself:  Using the gauge invariant field strengths, one
can readily check that the source-free equations of motion
are satisfied in the region $v \neq b$.  Of course, string charge
is not in fact conserved, and we therefore conclude that the membrane is 
configured just so as to annihilate exactly the amount of string charge
that enters
from infinity.  

Running the argument backwards, we could say
that we begin with zero string charge at the vertex and that
that quantity of string charge exiting to infinity is exactly
the amount produced by the membrane.  Thus, it is natural
to view the membrane as the primary source for the fields, and
the strings as a secondary consequence\footnote{Admittedly,
this viewpoint is adapted to the high symmetry of this
situation.  If one now adds to our solution the proper small flux
of strings coming in
along the paraboloid from infinity near $\psi =0$ and
exiting along the paraboloid near $\psi =\pi$ one should be able to, 
for example, obtain a solution in which the D4-brane charge
still vanishes at some point on the paraboloid, though this point
will no longer be at the vertex.  It would be interesting to know whether
such a solution breaks further supersymmetries. One would suspect that it does, since it clearly breaks rotational invariance and supersymmetries square to Killing vectors.}.
In this way, we can view the string ending on the monopole
as being equivalent to the degenerate limit of the paraboloidal 
membrane.  We may call this process the transmutation,
catalyzed by the presence of the monopole, of a membrane into a string. 

\section{The M $\rightarrow$ IIA reduction.}

\label{review}

We now review the construction of \cite{Aki}, which produces
a distribution of NS5-, D4-, and D6-brane charge in type IIA
supergravity.  We proceed quickly, as the present case is in
direct parallel with the lower dimensional example presented in
section (\ref{ex}).

Consider the metric and gauge field corresponding to a solution containing
a $M5$--brane with charge $q$\cite{gueven} ,
\begin{equation}
ds_{11}^{2} = f^{-1/3}(-dx_0^2 +  dx_1^2 + dx_2^2 + dx_3^2 + dx_7^2 +
dx_{10}^2) +
f^{2/3}(dx_4^2 +  dx_5^2 + dx_6^2 + dx_8^2 + dx_9^2 )  \ ,
\label{M5}
\end{equation}

\begin{equation}
A^{(11)}_{[6]} = \frac{f-1}{f} dx_0 \wedge dx_1 \wedge dx_2 \wedge dx_3
\wedge dx_7 \wedge 
dx_{10} \ \ ,
\label{gauge}
\end{equation}
where,
$$
f = 1 + \frac{q}{r^3} \ \ \ \  \mbox{and}\ \ \   r^2 = x_4^2 + x_5^2 +
x_6^2 + x_8^2 +x_9^2 \ \ .
$$

 The field strength associated to $A^{(11)}_{[6]}$ is the $7$--form
$F^{(11)}_{[7]}=dA^{(11)}_{[6]}$. Its dual is the $4$--form 
$F^{(11)}_{[4]}= {}^*\! F^{(11)}_{[7]} $  which is straightforwardly computed
from (\ref{M5}) and (\ref{gauge}),
\begin{eqnarray}
F^{(11)}_{[4]} &=&  
\frac{3q}{r^5}\left( dx_4\wedge dx_5\wedge dx_6\wedge [x_8\wedge dx_9 -
x_9\wedge dx_8]  +\frac{}{} \right. \nonumber \\
  & & \left.+\frac{}{}
  [x_4 dx_5\wedge dx_6 + x_5 dx_6\wedge dx_4 + x_6  dx_4\wedge dx_5 ] \wedge
  dx_8\wedge dx_9      
 \right) \ .
\label{dual}
\end{eqnarray}

At this point the topology and differential structure of the spacetime
are those of ${\bf R}^{1,10}$. However, we 
are interested in the topology and differential structure of
${\bf R}^{1,6}\times {\bf C}^2/{\bf Z_N}$,
which is obtained as follows.  Define as in the preceding section
 $W=x_7+ix_{10}$ and $V=x_8+ix_9$. The ${\bf Z_N}$ orbifold  group acts by
rotating simultaneously this two complex coordinates,{\it i.e.},
\begin{eqnarray*}
{\bf Z_N} : {\bf C}^2 &\longrightarrow& {\bf C}^2 \\
(V,W) &\longmapsto& e^{\frac{2\pi i}{N}}(V,W) \ \ .
\end{eqnarray*}
${\bf C}^2/{\bf Z_N}$ is the  $4$--dimensional manifold defined by
identifying points in ${\bf C}^2$ which are related by the action of
${\bf Z_N}$. Our spacetime manifold is topologically the product of
$R^{6,1}$ (in $x_0,\ldots  x_6$) and ${\bf C}^2/{\bf Z_N}$ (in
$x_7,\ldots, x_{10}$).  Clearly, all the fields are invariant under
rotations in the $(x_7,x_{10})$ and $(x_8,x_9)$ planes. 
Therefore defining $W=w e^{i\phi/N}$ and $V = v e^{i(\psi + \phi/N)}$,
for $\phi, \psi \in[0,2\pi)$, 
the vector field $\partial_\phi$ is a
Killing 
vector field. 
One dimensionally reduces along this vector field to obtain the desired
10-dimensional solutions.  

Recall that dimensional reduction to $10$--dimensional type IIA
supergravity identifies the different fields
as follows, where $z$ is a coordinate along the Killing orbits used
in the reduction:
  \begin{eqnarray}
  ds_{(11)}^2 &=& e^{-2\Phi/3}ds^2 + e^{4\Phi/3}(dz+A_{[1]})^2
\label{metricred} \ \ , \\
  F^{(11)}_{[4]} &=& \tilde F_{[4]} + F_{[3]}\wedge (dz + A_{[1]}) \ . \
\label{gaugered}
  \end{eqnarray}
Here, $\Phi$ is the usual dilaton field and $ds^2$ is the metric in the string
frame. It will be useful to define coordinates in analogy with 
(\ref{VWtransf}) through
\begin{equation}
V   = \rho \sin \frac{\theta}{2}e^{\psi+\phi/N} \ , \ \ \ \ \ \ \ \ \  
W =  \rho \cos \frac{\theta}{2}e^{\phi/N} \ . 
\label{vw}
\end{equation}
Note that in these coordinates, the action of ${\bf Z_N}$ takes the
convenient form 
$\phi\longmapsto \phi +2\pi$.  As it is standard in dimensional
reduction to use a Killing vector field with dimensions of inverse length, one
takes
$\partial_z=\partial_\phi /R_{11}$ for some fixed radius $R_{11}$.  To
present the dimensionally reduced field strengths it is convenient to first
introduce spherical coordinates in ($x_4,x_5,x_6,x_8,x_9$) defined by 
\begin{center}
$x_4 =  r \cos\alpha \ ,  \ \ \ \ \ \ x_5 =  r  \sin\alpha \cos\beta\ ,  \ \ \ \ \ \ 
x_6 =  r  \sin\alpha  \sin\beta \cos\gamma \  , $  \\
$x_8 =  r  \sin\alpha  \sin\beta \sin \gamma \cos(\psi+\phi/N) \ \  \ \ \  
x_9 = r \sin\alpha \sin\beta \sin \gamma  \sin(\psi+\phi/N)\ , $  \\
$x_7 = w \cos(\phi/N) \ , \ \ \ \ \ \ \ \ \ \   x_{10} = w \sin(\phi/N)  \ , $
\end{center}
for $0\le\alpha,\beta,\gamma<\pi$, $0\le \psi,\phi\ < 2\pi$ and $r, \ w$ real
and positive.  Taking $(r,\alpha,\beta,\gamma,\psi,w)$ to label the Killing
orbits leads to the same reduction as (\ref{vw}).
 Now, from (\ref{gaugered}) we obtain,
  \begin{eqnarray}
 \tilde F_{[4]} &=& -\frac{3qw^2\sin^5\alpha \sin^4\beta
\sin^3\gamma}{r^2\sin^2\alpha \sin^2\beta
\sin^2\gamma\left(1+\frac{q}{r^3}\right) + w^2}
  \ d\alpha\wedge d\beta\wedge d\gamma\wedge d\psi
 \label{F4} \ \ , \\
 F_{[3]} &=& -\frac{3q}{NR_{11}}\sin^3\alpha \sin^2\beta \sin\gamma \
d\alpha\wedge d\beta\wedge d\gamma
 \label{F3} \ \ ,
 \end{eqnarray}
 while the $1$--form is given by,
 \begin{equation}
 A_{[1]} = N R_{11}\frac{r^2\left(1+\frac{q}{r^3}\right)\sin^2\alpha
\sin^2\beta \sin^2\gamma}{w^2 + r^2\left(1+\frac{q}{r^3}\right)\sin^2\alpha
\sin^2\beta \sin^2\gamma} 
d\psi \ .
 \label{A1}
 \end{equation}

The ten-dimensional fields above are somewhat different from the fields
originally presented in \cite{Aki}.  In particular, our 3-form field
strength $F_{[3]}$ is non-zero as given by (\ref{F3}).  This difference
is due to certain subtleties associated with dimensional reduction and
Hodge duals in spacetimes with non-diagonal metrics that were not
taken fully into account in \cite{Aki}.  In particular, it is the
field strength $F_{[4]}^{(11)}$ that is related to the standard ten-dimensional
fields by (\ref{gaugered}), while the relation of its Hodge dual 7-form
field strength ($F^{(11)}_{[7]}$) to the
ten-dimensional fields is more complicated.

Note that (\ref{A1}) represents a
monopole located at the origin of the reduced ALE space.  Its world--volume
spans the entire ${\bf R}^{1,6}$ factor, which shows it to be
a $6$--dimensional magnetic object: a $D6$--brane of type IIA supergravity.
 As usual, the magnetic
potential shows a string--like singularity, extending from the origin to
infinity through the negative $Z$--axis of the reduced ALE space. This
corresponds to the points $w =0$ in the  above coordinates. As
in section \ref{ex}, we may again generalize to a family of solutions
constructed from a set of M5-branes oriented along the 5-space $(W, 
\overline{W}, x_1, x_2, x_3)$ and smeared over the circle $|V|=b$.

\section{Brane Mapping}
\label{mapping}

Having reviewed the basic construction of the solution in section 
\ref{review}, we need only check a few basic features in order to 
map out all of the branes and
show conclusively that it
can be interpreted in parallel with the discussion of
section \ref{ex}.  One distracting feature of the
present case is that the magnetic monopole associated with the
$F_{[2]}$ resulting
from the Kaluza-Klein reduction is no longer a point defect, but
an entire D6-brane.  This makes a discussion of charge
conservation more subtle as D4-brane charge (the analogue of
string charge in section \ref{ex}) is now measured
not just with a circle around the $\psi$ axis, but with
a 5-sphere which extends in certain directions along the 
D6-brane.  As a result, the corresponding 5-sphere can no longer
be deformed to a point without intersecting any charge.
What happens is that, when one tries to pull the 5-sphere down
into the lower half of the ALE space, one necessarily encounters
some part of the D6-brane located at some nonzero value of
$x_4, x_5,$ or $x_6$.  However, as we will see, this feature
is a mere distraction and does not significantly change the 
interpretation of the solution.

Let us first comment on charge conservation.
To clarify this point, note that from (\ref{gaugered}) the equation of
motion for $\tilde{F}_{[4]}$ away from any source is
\begin{equation}
\label{critEOM}
d \tilde F_{[4]} + F_{[2]} \wedge F_{[3]} =0.
\label{uu}
\end{equation}
Here we have used the fact that $dF^{(11)}_{[4]}=0$ outside the sources,
and  that $F_{[3]}$ is closed, which is also clear from (\ref{gaugered}).
When $4$--branes are present, equation (\ref{uu}) is written
\begin{equation}
d \tilde F_{[4]} + F_{[2]} \wedge F_{[3]} ={}^*\!  j_{4}
\end{equation} 
in analogy with
(\ref{charge}),
where now ${}^*\! j_4$ is the Hodge dual of the current associated with
$4$--brane charge. Taking the exterior derivative of this equation we find,
in analogy with (\ref{cnc}),
\begin{equation}
d {}^* \! j_4 = {}^*\!  j_6 \wedge F_{[3]} + F_{[2]} \wedge {}^*\!  j_5,
\label{cnc2}
\end{equation}
where  $j_6$ and $j_5$ are the currents associated to  D$6$-- and
NS$5$--branes respectively. This shows that the $4$--brane charge need not
be conserved in the presence of NS5--  and D6--branes, which is indeed
the case in which we are interested\footnote{This is, of course, only
a special case of a general subtlety involving charge conservation in
supergravity that arises whenever the gauge invariant field strength
satisfied an equation of motion of the form (\ref{critEOM}).
Thus, similar non-conservation effects hold for M2-brane charge in
the presence of M5-brane charge, for strings in the presence of various
D-branes, etc.}.

We now check that
all of the branes are in fact located on the paraboloid at
$v=b$.  To do so, recall
that for the original $11$--dimensional solution, we had
$dF^{(11)}_{[4]}={}^*\! j^{(11)}_5$. The general case $b\neq 0$ is
obtained\cite{Aki} by including a set of $M5$--branes smeared
along  $\partial_\phi$
at $|V|=b$ . In this way we keep the original symmetry along
$\partial_{\phi}$, and the
($11$--dimensional) Hodge dual of the current is,
 $$
 *\! j^{(11)}_5 = \frac{Q}{2\pi} \delta(x_4)\delta(x_5)\delta(x_6)\delta(v-b)
 dx_4 \wedge dx_5 \wedge dx_6 \wedge dv \wedge d\tilde{\psi} \ ,
 $$
 where $\tilde{\psi}=\psi +\phi/N$ is the angle in the $(x_8,x_9)$ plane.
{}From this expression we can easily get the equations of motion for the
$10$--dimensional fields,
\begin{eqnarray}
d{}^*\! F_{[3]} &=& {}^*\! j_5 = \frac{Q}{2\pi N R_{11}}
\delta(x_4)\delta(x_5)\delta(x_6)\delta(v-b)
 dx_4 \wedge dx_5 \wedge dx_6 \wedge dv \ , \label{j4}\\
d \tilde F_{[4]} + F_{[2]} \wedge F_{[3]} &=& {}^*\! j_4 = 
\frac{Q}{2\pi } \delta(x_4)\delta(x_5)\delta(x_6)\delta(v-b)
 dx_4 \wedge dx_5 \wedge dx_6 \wedge dv \wedge \left(d\psi -
\frac{A_{[1]}}{NR_{11}}\right) \ .
\nonumber
\end{eqnarray} 
In such coordinates $A_{[1]}$ takes the form,
\begin{equation}
A_{[1]} = \frac{NR_{11}}{1 + \frac{w^2}{f v^2}}d\psi  \ \ .
\label{AA1}
\end{equation}
{}From  (\ref{j4}) and (\ref{AA1}) we immediately see that there can be
$4$--brane charge only  on the surface of the paraboloid at $x_4=x_5=x_6=0$,
$v=b$.  Note that ${}^*\! j_4$  goes to zero at both $v=0$ and $w=0$. This
rules out the possibility of having a singularity at those points, where
the form $d\psi$ is not well defined.  The harmonic function $f$, which
also diverges only at the paraboloid, is given by
\begin{equation}
f=1 + \frac{Q}{16\pi^3}\int_0^{2\pi}
\frac{d\mu}{\left(s^2+|v-be^{i\mu}|^2\right)^{3/2}}  \ ,
\label{harmonic}
\end{equation}
where $s^2=x_4^2+x_5^2+x_6^2$. 
As a result, it is clear that all branes lie along the paraboloid
$v = b$.  

The field strengths of the $10$--dimensional theory can be obtained from
the $11$--dimensional solution which, in this generic case is,
\begin{eqnarray}
F^{(11)}_{[4]} &=&  \left( A v [x_4 dx_5\wedge dx_6 + x_5 dx_6\wedge dx_4 +
x_6  dx_4\wedge dx_5 ] \wedge  dv \right. \nonumber \\
&&-\left. B v^2 [ dx_4\wedge dx_5\wedge dx_6]\right) \wedge d(\psi +\phi/N)
\ ,
\label{f114b}
\end{eqnarray}
with
\begin{eqnarray}
A&=& \frac{-3Q}{16\pi^3}\int_0^{2\pi} \frac{d\mu}{\left(s^2+v^2+b^2-
2bv\cos\mu \right)^{5/2}} \ , \\ 
B&=& \frac{-3Q}{16\pi^3}\int_0^{2\pi}
\frac{(1-\frac{b}{v}\cos\mu)d\mu}{\left(s^2+v^2+b^2- 2bv\cos\mu
\right)^{5/2}} \ .
\end{eqnarray}

Now, for non-zero $b$, we may easily work around
the distraction of the D6-brane in
interpreting the D4-brane charge distribution.
By symmetry, any D4-branes on the paraboloid must enter
in a rotationally invariant fashion along the opening at
infinity and run along surfaces of constant $\psi$.
Consider then the subsurface $S$ at some constant $\psi$, 
$v=b$, and $x_4, x_5, x_6 =0$ which would describe the
world-surface of such a D4-brane.  Note that we can surround it
with a small 5-sphere that can be deformed from infinity
down to the vertex of the paraboloid without intersecting any
charge (and, in particular, without intersecting the D6-brane
at the origin of the ALE space).  
Two-dimensional projections of such spheres lie on the boundaries of the tubes
shown in Fig. 3 below.
Thus, for $b\neq 0$,
we may discuss conservation of D4-brane charge in direct parallel
with section \ref{ex}.  Since there are no branes exiting the
vertex of the paraboloid along the $\theta=0,\pi$ axes, the
D4-brane charge at the vertex must vanish by rotational invariance, as can
be checked directly
 from (\ref{j4}). It follows
that all D4-brane charge may be thought of as generated by the
NS5-brane in the background field of the D6-brane through the
charge non-conservation laws of type IIA supergravity.
\setcounter{figure}{2}
\EPSFIGURE[ht]{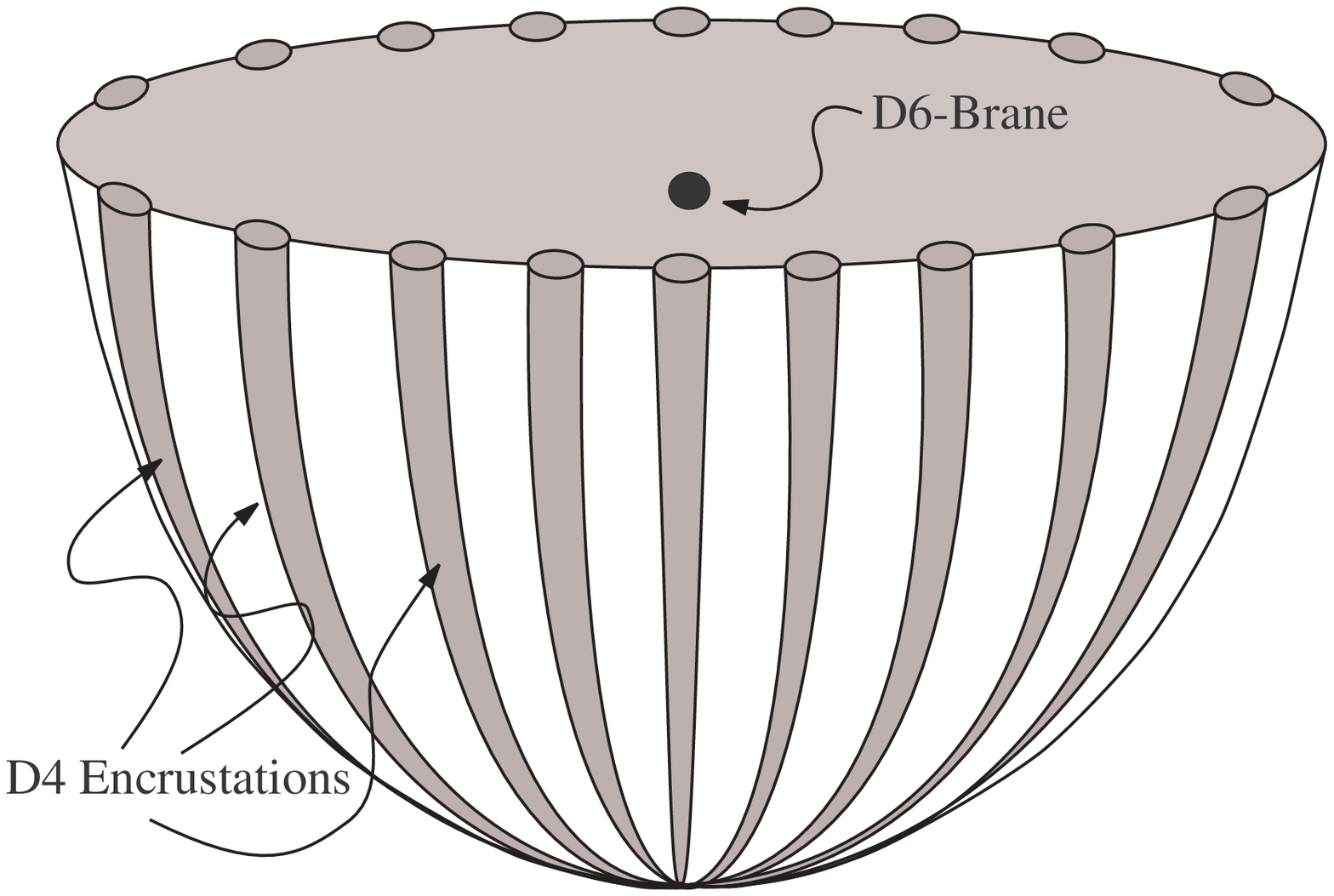,width=10cm}{The D4--charge goes to a constant value for large values of
$w$. For $w=0$, at the vertex of the paraboloid, the charge goes to zero
which is possible due to the non conservation of this charge. The
individual D4--brane  sources are distributed along the paraboloid at
values of constant $\psi$ and can be seen as encrustations on the 5--brane.}

We would now like to address the issue of the finite
$F_{[3]}$ field strength created by the pure multipole configuration
of the NS5-brane in the $b \rightarrow 0$ limit.
The interesting difference between this and, say, point dipoles
in familiar Maxwell electrodynamics is that in the familiar Maxwell
case such non-trivial pure dipole fields arise only when an
infinite amount of charge is present.  It is thus of interest
to quickly check that the total amount of NS5-brane does not
diverge as $b \rightarrow 0$, and that it is in fact independent
of $b$. This can be seen from (\ref{j4}), but we will show an independent 
computation in order to further illuminate the geometry of the system.

Define a cube letting $-L \le x_4,x_5,x_6 \le L $ with $v=0$ and $w=w_0$
fixed. A second cube is defined in the same way but with $v=v_0 > b$.  We
finish closing  our $3$--manifold  with the surface defined by the
interval $0\le v$ and the faces of the previous cube in $(x_4,x_5,x_6)$.

This $3$ surface clearly encloses the $5$--brane, and we can easily compute
the integral of $F_{[3]}$ over it. To do so first note that, due to
(\ref{f114b}), the integral over the cube at $v=0$ is zero. The integral
over the  surface joining the two cubes goes to zero in the limit
$L\rightarrow\infty$. To see this, let us integrate over one  of this
surfaces, say, $x_4=L$,
\begin{equation}
\int_{v=0}^{v_0}\int_{x_5=-L}^L \int_{x_6=-L}^L A L dx_5dx_6 dv \ .
\end{equation}
As $L\rightarrow\infty$ this integral vanishes as $v_0^2/L^2$.
The  charge of the $NS5$--brane is given by the remaining integral,
therefore, 
\begin{equation}
Q_5=\frac{1}{NR_{11}}\int_{-\infty}^{\infty}dx_4
\int_{-\infty}^{\infty}dx_5 \int_{-\infty}^{\infty}dx_6
\frac{v_0^2 B}{(x_4^2+x_5^2+x_6^2+v_0^2)^{5/2}} = \frac{Q}{2\pi NR_{11}}  \
\ ,
\label{q5}
\end{equation} 
which is indeed finite and independent of $b$.

\section{Discussion}
\label{disc}

We have studied a family of type IIA supergravity
solutions obtained by the reduction of a family of 11-dimensional
solutions.  One member of this family describes
a pure multipole configuration
of NS5-branes stabilized by the presence of a D6-brane.  
In this pure-multipole limit, the NS5-brane in fact
becomes a D4-brane ending on the D6-brane.  By varying
a parameter $b$ from $\infty$ to zero, we obtain a family
of BPS solutions 
in which an NS5-brane moves in from infinity, curves around
the D6-brane, and degenerates into the D4-brane ending on the
D6-brane.  This sort of brane transmutation arises due to
the lack of a conservation law for certain types of charge
in type IIA supergravity, so that an NS5-brane in the background
field of a D6-brane can act as a source of D4-brane charge.
It is clear from the 11-dimensional description
that all members of this family preserve the same supersymmetries, 
so that our family in fact forms a moduli space.  By the usual
adiabatic arguments, this moduli space can therefore be 
associated with a dynamical process in which a slowly moving NS5-brane
approaches a D6-brane, wraps around it, and transmutes itself into
a D4-brane.   It is reassuring that our results bear a close
resemblance to those of \cite{Im}, which study a curved test brane
(a D5-brane, in that case) in the background of another brane (a D3)
using Born-Infeld techniques.

Although
we have studied only the near-core solution in explicit form, 
we expect a corresponding asymptotically flat solution with similar
properties.    This is indicated, for example, by the results of
\cite{Cal} which extend the Born-Infeld solution of \cite{Im} into
the asymptotically flat region. 
One might at first expect that the NS5-brane
remains flat far from the D6-brane, in which case the asymptotically
flat analogue of our $b=0$ solution would resemble the diagram below.
This is a clear analogue of the brane-creation processes discussed
by Hanany and Witten \cite{HW} and others \cite{Dan,Bachas,Bergman}.
As a result, the moduli space mentioned above provides a dynamical
supergravity description of such phenomena.

\begin{center}
 \epsfxsize=8cm
 \leavevmode
 \epsfbox{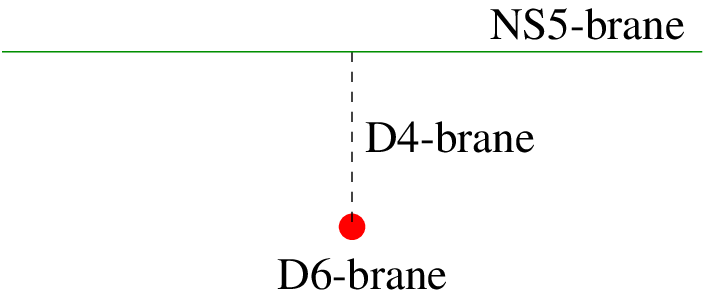}
\end{center}
\nopagebreak
\centerline{\small {\bf Figure 4:}  A first guess for the asymptotically flat case.}

\noindent

We recall, however, that when a given brane (our D4-brane)
ends on a brane of one dimension higher (the NS5-brane), the larger
brane takes the shape of a logarithmic curve and does not quite become
flat at infinity\cite{bend}.  Thus, the figure above will be somewhat modified.
In any case, we expect a limit in which the flatter parts of the NS5-brane
move up the page and off to infinity so that
the total NS5-brane charge vanishes as measured by any
Gaussian surface in the spacetime.  
This limiting solution represents a D4-brane ending on a D6-brane in
the presence of a nonzero Neveu-Schwarz field strength.

As this solution has three translational symmetries, it
is straightforward
to T-dualize up to three times, obtaining solutions with multipole
NS5-branes representing D3-branes ending on D5-branes,
D2-branes ending on D4-branes,  or D1-branes ending on D3-branes.  
Either of the
type IIB solutions are readily S-dualized as well, and this last
one then yields a smeared solution containing fundamental strings
ending on a
D3-brane stabilizing a pure multipole configuration of D5-brane.     

Returning to our near-core solutions, one
issue that we should pin down more carefully is whether
pure-multipole configurations of NS5-brane charge are in
fact generically associated with the ending of a D4-brane
on a D6-brane or whether there exists another solution in which only
the fields usually associated with D4- and D6-branes are
excited.  That the presence of a nonzero Neveu-Schwarz
field {\it is} generic can be argued
by charge conservation.  As discussed in section \ref{ex}, 
the presence of a D6-brane is not by itself enough
to foil conservation of D4-brane charge.  Breaking charge
conservation necessarily requires the presence of a Ramond-Ramond
field as well.   However, one might seek a solution in which D4-brane
charge {\it is} conserved but merely flows away along the D6-brane
in the $x_4,x_5,x_6$ directions.
Indeed, a striking feature of our solution is that no D4-brane charge
leaks into the D6-brane at all.

We therefore turn to a second argument that the NS-field is generic. 
Let us T-dualize
our solution 3 times (to obtain a D-string ending on a D3-brane)
and then S-dualize to obtain a fundamental string ending on
a D3-brane.  T-duality will leave our tightly wrapped
NS5-brane an NS5-brane, and S-duality will turn it into
a D5-brane.  Thus, if an NS5-brane is always associated
with a D4-bane ending on a D6-brane, we would expect
a fundamental string ending on a D3-brane to be associated with
a `pure-multipole' D5-brane.  This brane-ending configuration
was investigated in the appendix of \cite{dj} using 
perturbative techniques and the Dirac-Born-Infeld action for the
branes.  A look at the Ramond-Ramond fields
found there does indeed show the presence of a 
magnetic fields of the sort that would be associated with a 
certain pure-multipole configuration of D5-branes.  Note, however, that
the charge non-conservation effect could not be seen in \cite{dj}
as the fields were computed there only to first order in the
charges, while we see from (\ref{cnc}) that charge non-conservation
is a second order effect.  Similarly, charge non-conservation will not
be seen in other lowest order perturbative calculations.

Finally, a third argument that the solution studied above
is in fact `the' solution for D4-branes ending on D6-branes
is obtained by considering what happens when we add a
second paraboloid of NS5-branes enclosing the negative axis.
This is most easily studied in the simplified example
of section \ref{ex}, but may be done in the full 11-dimensional supergravity
case as well.  One begins with the spacetime
considered above having branes at $|V|=b$ oriented along the $W$
plane. 
One then inserts an additional
set of branes oriented
along the $V$ plane and located at $|W|=b$.  In the limit 
$b \rightarrow 0$, we obtain two orthogonally intersecting
branes at $V=0$ and $W=0$.  After Kaluza-Klein reduction, we find strings
along the positive and negative axes.  It is readily checked that
the Neveu-Schwarz field strength vanishes
when the signs are chosen such that these form a continuous
D4-brane passing through the D6-brane.  Thus, assembling two
of our half-branes does in fact make a familiar Neveu-Schwarz-free
D4-brane.

This last argument allows us to easily check that
the general solution from section \ref{review}, 
which contains a mixture of D4-, NS5-, and
D6-brane charge, is in fact a BPS configuration.
The point is that the 11-dimensional
configuration of two M5-branes at $V=0$ and $W=0$ with signs as above
is a typical
`branes at angles' solution (see \cite{angles}) preserving
1/4 of the supersymmetries.  Now, since the type IIA configuration with a
D4-brane orthogonally intersecting a D6-brane is also a 1/4 BPS
solution, we see that no supersymmetries of the branes at angles
solution are broken in the reduction process.  Since all of these
supersymmetries are also present in the case of a single set of branes
oriented along the $W$-plane  (located at either $V=0$ or $|V| = b$),
we can indeed be sure that our type IIA
solutions represent a moduli space of BPS configurations
preserving the same 1/4 of the supersymmetries.

Having concluded that the $b=0$ case does indeed represent
`the' solution for a D4-brane ending on a D6-brane, we deduce
that another qualitative
difference should arise between the near-core solutions studied above
and the full asymptotically flat
solutions.  In our near core solutions, the field strengths
$F_{[3]}$ and $\tilde{F}_{[4]}$ associated with NS5- and D4-brane
charge are of comparable magnitude.  In fact, 
in an appropriate gauge we have just $\tilde{F}_{[4]} =
F_{[3]} \wedge (d \psi/NR_{11}-A_{[1]})$, where $NR_{11}$ is roughly the tension of the
D6-brane.  In the full asymptotically flat
solution for $b=0$, we would expect the $F_{[3]}$ field to fall off
faster than the $\tilde{F}_{[4]}$ field, reducing to that of a pure
D4-brane as we move along the D4-brane and away from the 
D6-brane.  Something like this is seen, for example, in
the perturbative calculation in \cite{dj}
of the fields generated by a fundamental string ending
on a D3-brane.  It appears that the degeneracy seen in the
near-core solutions is a consequence of the fact that the near-core region of
the Kaluza-Klein monopole has an $SO(4)$ rotational symmetry, while
the full solution breaks this to $SO(3) \times U(1)$.  

A final natural question to ask is whether all 
brane ending phenomena in which a `half-brane' ends on
a `terminal brane' arise in this way.  That is, can they all arise
as a case of brane transmutation, with the
resulting half-branes being formed as some
larger brane wraps itself tightly around the terminal brane?
If this is the case, then all brane-ending solutions must resemble
the one above, in which the charge of the half-brane is in fact absorbed
by the terminal brane and does not flow along the terminal brane to 
infinity.
For a fundamental string ending on a D0-brane, it is clear
that this is the case, and the
Born-Infeld results of \cite{Im,Cal} lead to a similar description
of strings stretched between D3-branes and D5-branes.
While we cannot reach a definite general conclusion here, we note
that the charge non-conservation
effect arises from Chern-Simons-like couplings and that 
an analysis of such couplings as in \cite{surgery}
does make an affirmative answer seem likely. We expect such
considerations to provide interesting
examples for further study.

\acknowledgments

The authors would like to thank Jerome Gauntlett, 
Akikazu Hashimoto,  Amanda Peet and Arkady Tseytlin
for useful discussions.  This work was supported in part by
NSF grant PHY94-07194 to the ITP (Santa Barbara),
NSF grant PHY97-22362 to Syracuse University, 
the Alfred P. Sloan foundation, and by funds from Syracuse 
University.


\end{document}